\documentclass[10pt,twocolumn,aps,prl,groupedaddress]{revtex4-1}

\usepackage{amsmath}
\usepackage{amssymb}
\usepackage{graphicx}
\usepackage{soul}

\def\bea{\begin{eqnarray}}
\def\eea{\end{eqnarray}}
\def\ben{\begin{equation}}
\def\een{\end{equation}}
\def\benu{\begin{enumerate}}
\def\enu{\end{enumerate}}

\def\lsim {\ifmmode {\buildrel<\over\sim}}

\def\sss{\scriptscriptstyle\rm}

\def\1var{(\bx_1...\bx\N)}

\def\half{\frac{1}{2}}

\def\br{{\bf r}}
\def\bR{{\bf R}}

\def\b1{{\bf 1}}
\def\bx{{x}}

\def\x{_{\sss X}}

\def\xc{_{\sss XC}}
\def\Hx{_{\sss HX}}

\def\N{_{\sss N}}
\def\H{_{\sss H}}

\def\sph_int{ {\int d^3 r}}

\def\infintd3r{ \int_{-\infty}^\infty d^3r\,}
\def\intd3r{ \int d^3r\,}

\def\laplace1d{\frac{d^2}{dx^2}}
\def\plaplace1d{\frac{d^2}{d{x'}^2}}

\def\padr2{\frac{\partial^2}{\partial r^2}}

\def\N{{\cal N}}

\def\b{{\beta}}

\begin{document}


\title{Density Functional Resonance Theory of Unbound Electronic Systems}


\author{Daniel L. Whitenack}
\email[]{dwhitena@purdue.edu}
\homepage[]{http://www.purdue.edu/dft}
\affiliation{Department of Physics, Purdue University, 525 Northwestern Avenue, West Lafayette, IN 47907, USA}
\author{Adam Wasserman}
\email[]{awasser@purdue.edu}
\affiliation{Department of Chemistry, Purdue University, 560 Oval Drive, West Lafayette, IN 47907, USA}
\affiliation{Department of Physics, Purdue University, 525 Northwestern Avenue, West Lafayette, IN 47907, USA}


\date{\today}

\begin{abstract}
Density Functional Resonance Theory (DFRT) is a complex-scaled version of ground-state Density Functional Theory (DFT) 
that allows one to calculate the resonance energies and lifetimes of metastable anions. 
In this formalism, the exact energy and lifetime 
of the lowest-energy resonance of unbound systems is encoded into a complex ``density'' that can be obtained via complex-coordinate scaling. 
This complex density is used as the primary variable in a DFRT calculation just as the ground-state density would be used 
as the primary variable in DFT.  As in DFT, there exists a mapping of the $N$-electron interacting system 
to a Kohn-Sham system of $N$ non-interacting particles in DFRT.  This mapping facilitates self consistent calculations 
with an initial guess for the complex density, as illustrated with an exactly-solvable model system. Whereas DFRT yields in principle the {\em exact} resonance energy and lifetime of the 
interacting system, we find that neglecting the complex-correlation contribution leads to errors of similar magnitude to those of standard scattering close-coupling calculations under the bound-state approximation.
\end{abstract}


\maketitle


Density Functional Theory (DFT) \cite{hohenberg, kohn, parr} provides one of the most accurate and reliable 
methods to calculate the ground-state electronic properties of molecules, clusters, and materials from first principles. It is one of the workhorses of computational quantum chemistry \cite{Martin}. In addition, DFT's time-dependent extension (TDDFT) \cite{RG84} can now be applied to a wealth of excited-state and time-dependent properties in both linear and non-linear regimes \cite{TDDFT}. When the $N$-electron system of interest has no bound ground state, however, neither DFT nor TDDFT can be applied in a straightforward way. A correct DFT calculation converges to the true ground state by ionizing the system, thus leaving no reliable starting point for a subsequent TDDFT calculation on the $N$-electron system. In practice, a finite simulation box or basis set can make the system artificially bound \cite{LFB10,KSB11}, but information about the relevant lifetimes is lost in the process.

We address here this fundamental limitation of ground-state DFT, and propose a solution.
 

Consider a system of $N$ interacting electrons in an external potential $\tilde{v}(\br)$, with ground-state density $\tilde{n}(\br)$. The potential is set to be everywhere positive and go to a positive constant $C$ as $|\br|\to\infty$. The ground-state energy is $\tilde{E}>0$. We start by asking how the gound state density changes when a smooth step is added to $\tilde{v}(\br)$ at a radius $|\bR|$ that is larger than the range of $\tilde{v}(\br)$. The step is such that the new potential $v(\br)$ coincides with $\tilde{v}(\br)$ for $|\br|<|\bR|$ but goes to zero at infinity. Since $\tilde{v}(\br)$ is everywhere positive, all $N$ electrons tunnel out and $v(\br)$ supports no bound states.  The correct ground state energy is now $E = 0$, and the new density $n(\br)$ is delocalized through all space. In practical calculations, however, $v(\br)$ and $\tilde{v}(\br)$ cannot be distinguished if $|\bR|$ is beyond the size of the simulation box. The result provided by ground-state DFT using the exact exchange-correlation functional is not $E$, but $\tilde{E}>0$, and the density 
obtained is $\tilde{n}(\br)$ as if the system were bound. Even when the simulation box is large enough to include the steps, use of a finite basis-set of localized functions will artificially bind all electrons. Clearly, such calculations do not provide approximations to the true ground-state energy and density of $v(\br)$, but to those of its lowest-energy resonance (LER). 

The purpose of this letter is to establish an analog of KS-DFT that provides the in-principle exact LER-density along with its energy and lifetime for any finite $|\bR|$. As $|\bR|\to\infty$, the results coincide with those of standard KS-DFT.  For higher-energy resonances, TDDFT is needed as a matter of principle~\cite{KM09,FB06,*FB06b,*FB09}. 

First, we note that as $|\bR|\to\infty$, the complex density $n_\theta(\br)$ associated with the LER of
\ben
\hat{H}_{v}=\hat{T}+\hat{V}_{ee}+\int d\br \hat{n}(\br) v(\br)~~,
\label{e:H}
\een
becomes equal to the complex density $\tilde{n}_\theta(\br)$ associated to 
$\tilde{v}(\br e^{i\theta})$. In Eq.~\ref{e:H},
$\hat{T}=-\half\sum_{i=1}^N\nabla_i^2$ is the kinetic energy operator, $\hat{V}_{ee}=\sum_{i,j\neq i}^N|\br_i-\br_j|^{-1}$ is the electron-electron interaction, and 
$\hat{n}(\br)=\sum_{i=1}^N \delta(\br-\hat{\br}_i)$ is the density operator. (Atomic units are used throughout). 
To find $n_\theta(\br)$, 
we complex-scale $\hat{H}_{v}$ by multiplying all electron coordinates by the phase factor $e^{i\theta}$, diagonalize the 
resulting non-hermitian operator $\hat{H}_{v}^\theta$, and calculate the bi-expectation value of $\hat{n}(\br)$ as:
\ben
n_\theta(\br)=\langle \Psi_\theta^L|\hat{n}(\br)|\Psi_\theta^R\rangle~~,
\label{eq:den}
\een
where $|\Psi_\theta^R\rangle$ and $\langle \Psi_\theta^L|$ are the right and left eigenstates corresponding 
to the complex eigenvalue of $\hat{H}_{v}^\theta$ that has the smallest positive real part among all eigenvalues in the 
non-rotating spectrum of $\hat{H}_{v}^\theta$. For a detailed review of this technique and related methods in non-hermitian 
Quantum Mechanics, see ref.~\cite{WW10, M11}. The computational cost of this prescription scales exponentially with the number of particles.
Since $n_\theta(\br)\to \tilde{n}_\theta(\br)$ as $|\bR|\to\infty$, 
and since there is a one-to-one correspondence between 
$n_\theta(\br)$ and $v(\br e^{i\theta})$ \cite{E06,WM07}, the complex energy of the LER $E_\theta[n_\theta]$ goes to $\tilde{E}$ 
(not $E$), as $|\bR|\to\infty$. Its lifetime ${\cal L}$ is given by $(-2{\rm Im}(E_\theta))^{-1}$, and for any finite $|\bR|$,
\ben
E_\theta [n_\theta]={\cal E}[n_\theta]-\frac{i}{2}{\cal L}^{-1}[n_\theta]~~,
\een
where the resonance energy ${\cal E}$ tends to $\tilde{E}$ as $|\bR|\to\infty$.

To build a complex analog of 
Kohn-Sham DFT using $n_\theta(\br)$ 
as the basic variable, we first map the system of interacting electrons whose LER density is $n_\theta(\br)$ to one of $N$ particles 
moving independently in a complex ``Kohn-Sham'' potential $v_s^\theta(\br)$ defined such that its $N$ occupied complex orbitals 
$\{\phi_i^\theta(\br)\}$ yield the interacting LER-density via 
$n_\theta(\br)=\sum_{i=1}^N \langle\phi_i^{\theta,L}|\hat{n}(\br)|\phi_i^{\theta,R}\rangle$. The complex Kohn-Sham equations are:
\ben
\left(\begin{array}{cc}
\hat{h}_1 - \varepsilon_i & -\hat{h}_2 - 2\tau_i^{-1}\\
\hat{h}_2 + 2\tau_i^{-1} & \hat{h}_1 - \varepsilon_i\\
      \end{array}\right)
\left(\begin{array}{c}
{\mbox{Re}(\phi_i^\theta)}\\
{\mbox{Im}(\phi_i^\theta)}\\
      \end{array}\right)
= 0
~~,
\label{e:KS}
\een
where $\hat{h}_1 = -\half\cos(2\theta)\nabla^2 + \mbox{Re}(v_s^\theta(\br))$, and $\hat{h}_2=\half\sin(2\theta)\nabla^2+ \mbox{Im}(v_s^\theta(\br))$. The set of $\{\varepsilon_i\}$ and $\{\tau_i\}$ provide the orbital resonance energies and lifetimes of the
Kohn-Sham particles.  

Second, we write $E_\theta[n_\theta]$ as:
\begin{eqnarray}
E_\theta[n_\theta] &=& T_s^{\theta}[n_\theta]+\int d\br \  n_\theta(\br) v(\br e^{i\theta}) \nonumber \\
 & & + E\H^{\theta}[n_\theta] + E\xc^{\theta}[n_\theta] 
\label{e:hypothesis}
\end{eqnarray}
in analogy to standard KS-DFT, and require: $T_s^\theta[n_\theta]=e^{-2i\theta} T_s[n_\theta]$  and $E\H^\theta[n_\theta]=e^{-i\theta}E\H[n_\theta]$, 
where $T_s[n_\theta]$ and $E\H[n_\theta]$ are the standard non-interacting
kinetic energy and Hartree functionals evaluated at the complex densities. Eq.~\ref{e:hypothesis} then defines $E\xc^\theta[n_\theta]$.
The complex variational principle \cite{M11} along with the assumption that the orbitals used to construct the density can be expanded in an orthonormal basis leads to the Euler-Lagrange equation:
\ben
\frac{\delta E_\theta[n_\theta]}{\delta n_\theta}-\mu\int d\br n_\theta(\br) = 0~~.
\een 
Performing the variarion in Eq.~\ref{e:hypothesis} and comparing with Eq.~\ref{e:KS} leads to an expression for the Kohn-Sham potential that is again analogous to that of standard KS-DFT:
\ben
v_s^\theta(\br)=v(\br e^{i\theta})+e^{-i\theta}v\H[n_\theta](\br)+v\xc^\theta[n_\theta](\br)~~,
\label{eqn:vs}
\een
where $v\xc^\theta[n_\theta](\br)=\delta E\xc^\theta[n_\theta]/\delta n_\theta(\br)|_{\rm LER}$.

\begin{figure}
\scalebox{0.265}{\includegraphics{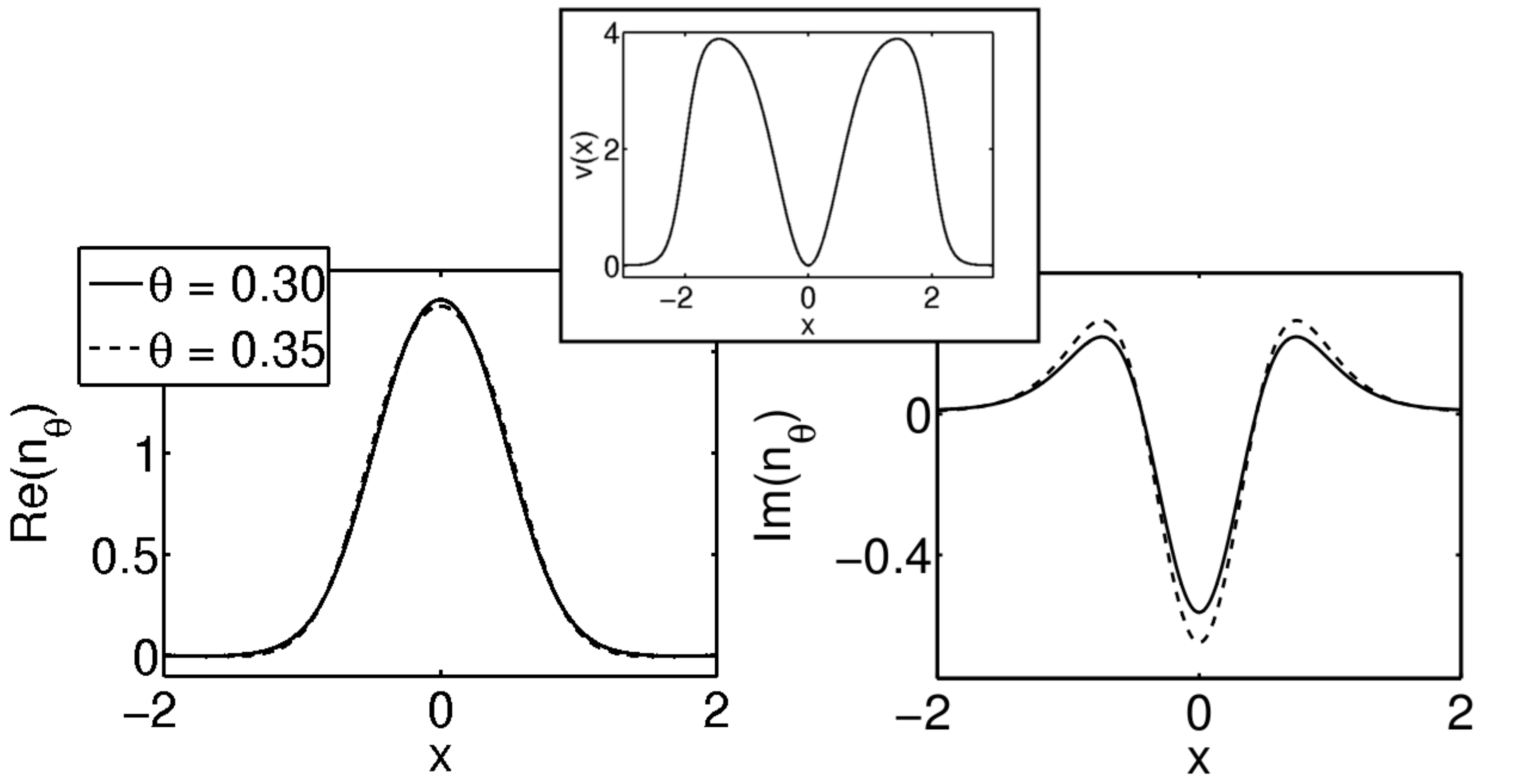}}
\caption{\label{fig:dendiff} Different exact 2-electron complex densities when using different scaling angles.  The model potential, $v(x)$, used in this study is also shown ($a=4$, $b=0.5$, $c=4$, and $d=2$).}
\end{figure}

The simplest case where all essential aspects of this formalism can be illustrated is a system of two interacting electrons moving in a one-dimensional
potential such as the one depicted in the inset of Fig.~\ref{fig:dendiff}. We study a Hamiltonian where the electrons interact via a soft-Coulomb potential of strength $\lambda$:
\ben
\hat{H}=\sum_{i=1}^2\left[-\half\frac{d^2}{dx_i^2}+v(x_i)\right]+\frac{\lambda}{\sqrt{1+(x_1-x_2)^2}}~~,
\label{e:Hamiltonian}
\een
using
$v(x) = a\left[\sum\limits^{2}_{j=1} \left(1+e^{-2 c (x + (-1)^jd)}\right)^{-1}- e^{-\frac{x^2}{b}} \right]$. 
Its parent potential $\tilde{v}(x)=a(1-e^{-x^2/b})$ goes to $a$ as $x\to\pm\infty$, but $v(x)$ goes down 
to zero at $x\sim \pm d$.

\begin{figure}  
\scalebox{0.265}{\includegraphics{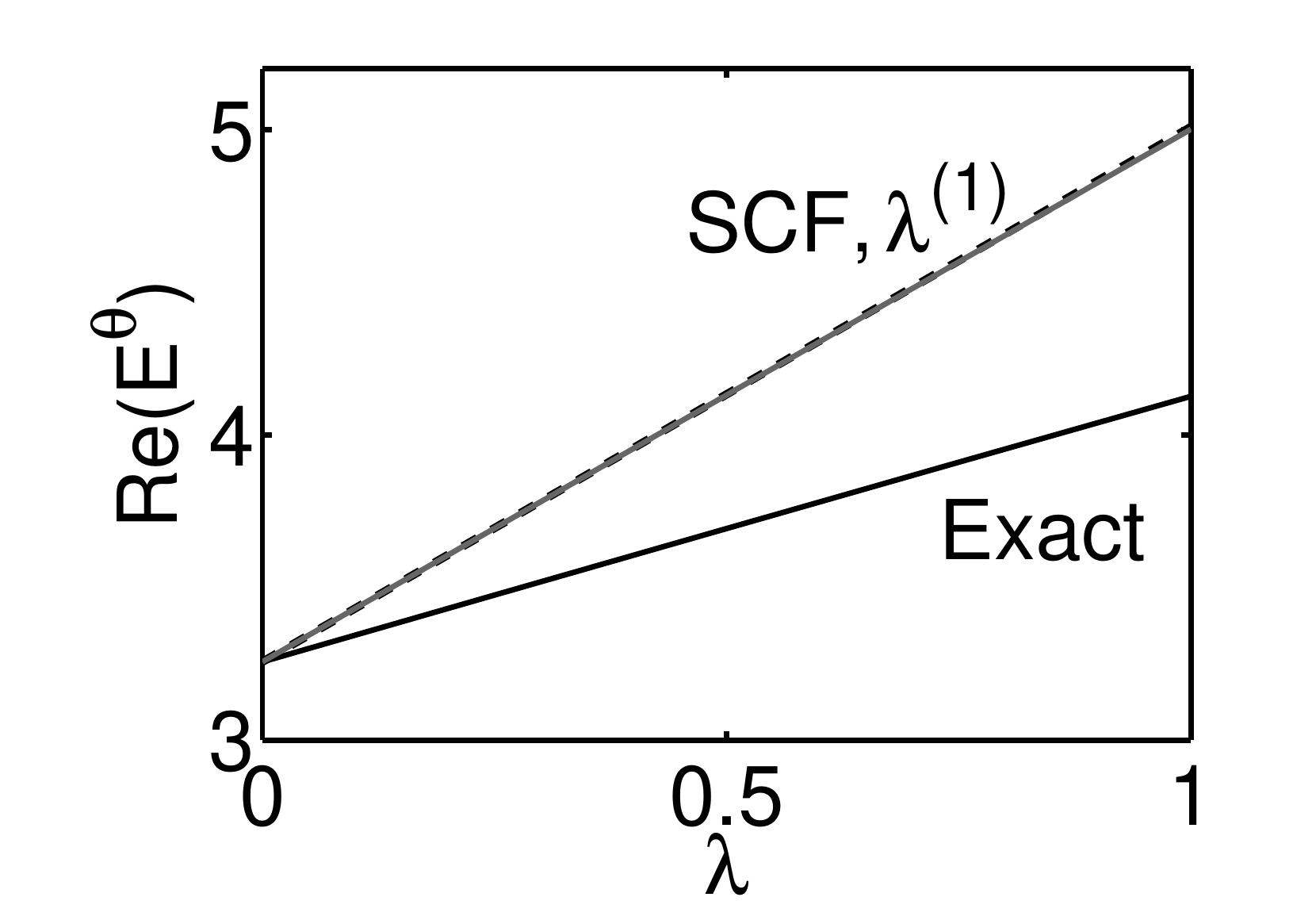}}
\hspace{-15pt}
\scalebox{0.265}{\includegraphics{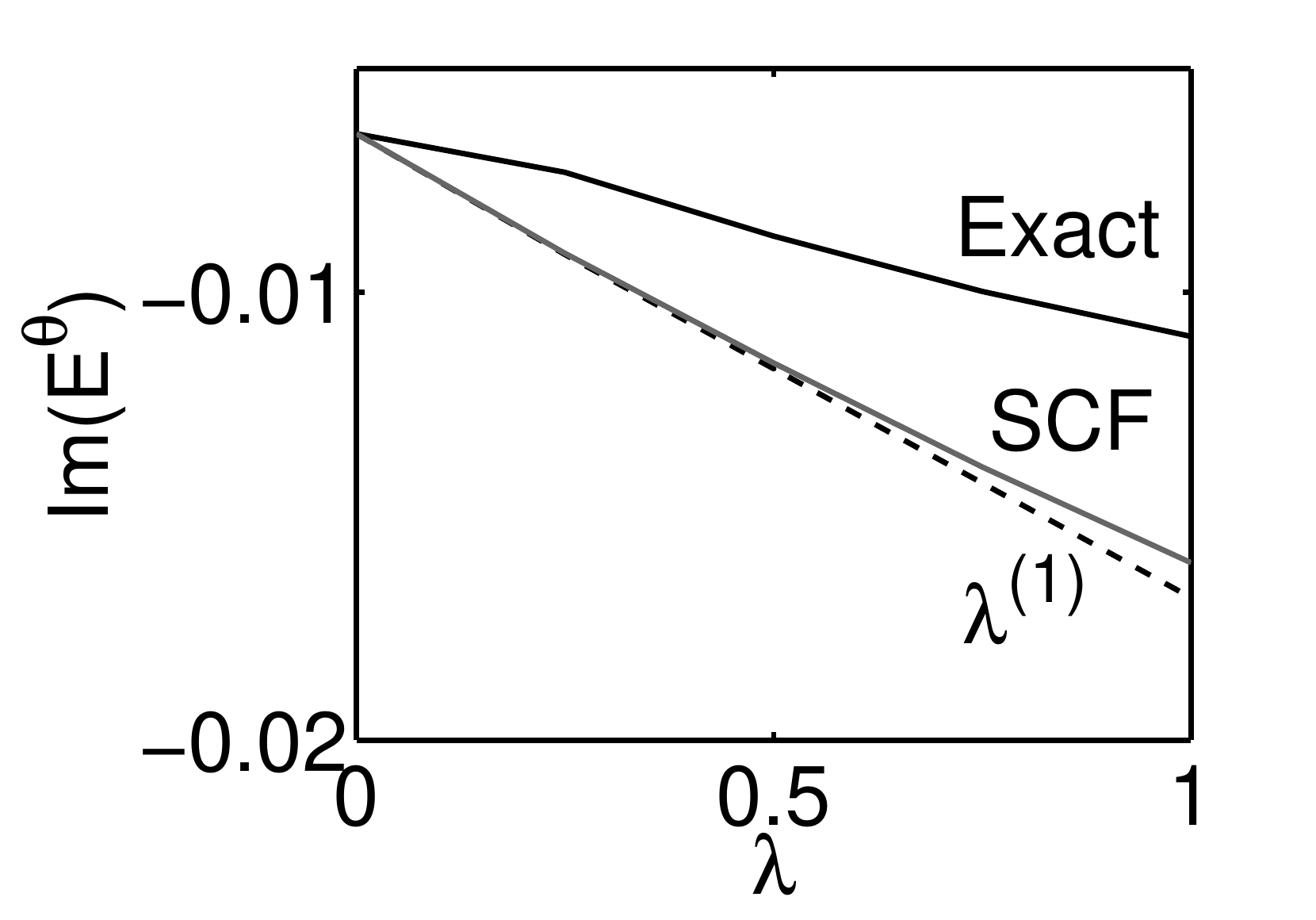}}
\caption{\label{fig:engs}The real and imaginary parts of $E_\theta$ in the model Hamiltonian of Eq.~\ref{e:Hamiltonian} calculated exactly with complex 
scaling (thick solid), a first order correction to the non-interacting energy (dashed), and our DFRT exchange-only self-consistent method (thin solid). 
($a=4$, $b=0.5$, $c=4$, and $d=2$)}
\end{figure}

{\underline {\em Exact solution via 2-electron wavefunction}}:
The complex-scaled Hamiltonian $\hat{H}_\theta=\hat{H}(\{x_i\}\to\{x_ie^{i\theta}\})$ was diagonalized with the Fourier Grid Hamiltonian (FGH) \cite{chu} and Finite Difference Methods. The numerically exact $n_\theta(x)$ was calculated via Eq.~\ref{eq:den}. The complex density $n_\theta(x)$ depends on the value of $\theta$ (see Fig.~\ref{fig:dendiff}), but for a large enough number 
of grid points the energy does not.  In the complex-scaling method the resonance energies are precisely those that remain stationary as $\theta$ changes~\cite{M11}. Fig.~\ref{fig:engs} shows the energy for $0<\lambda<1$.

\begin{figure}
\scalebox{0.265}{\includegraphics{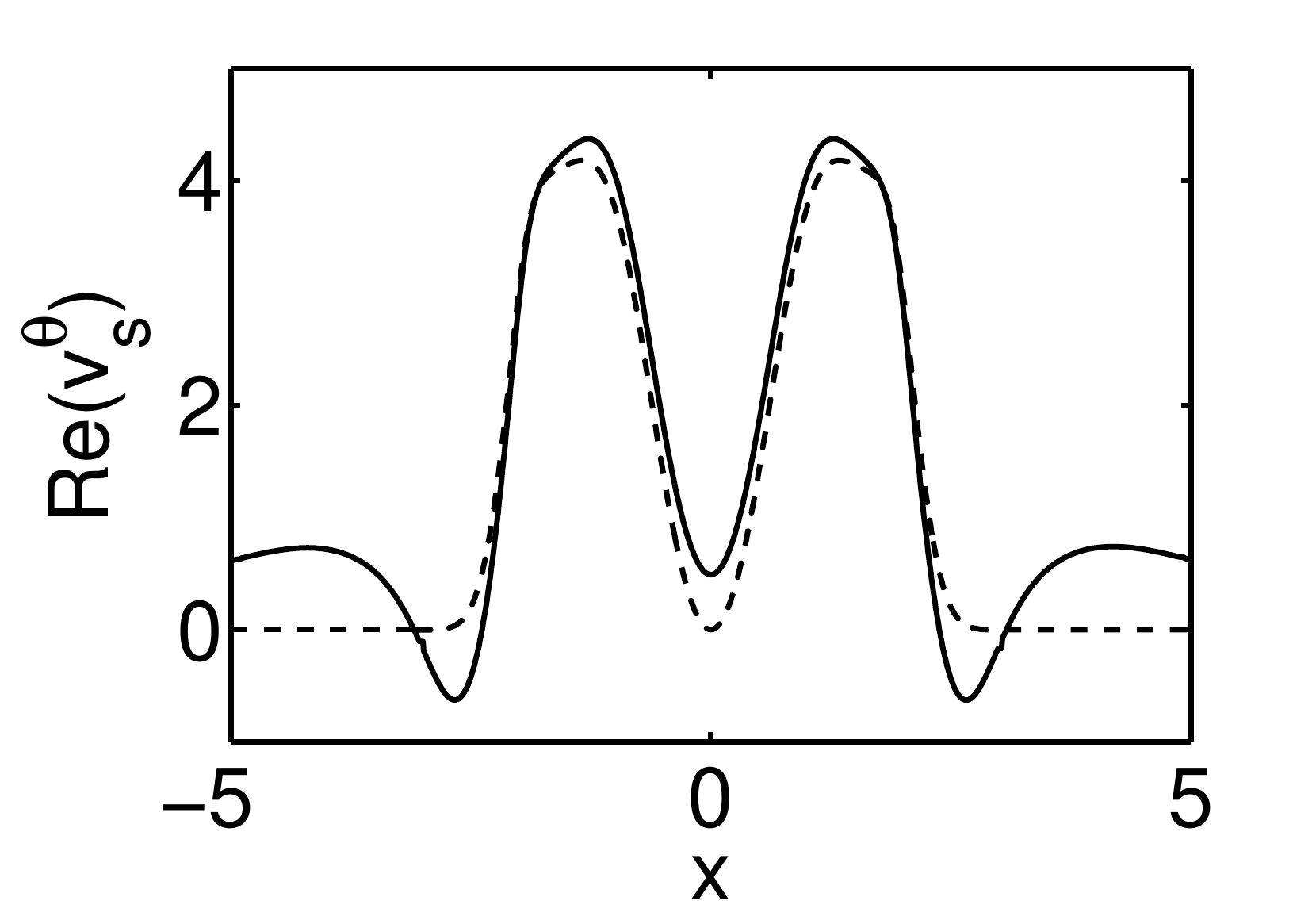}}
\hspace{-15pt}
\scalebox{0.265}{\includegraphics{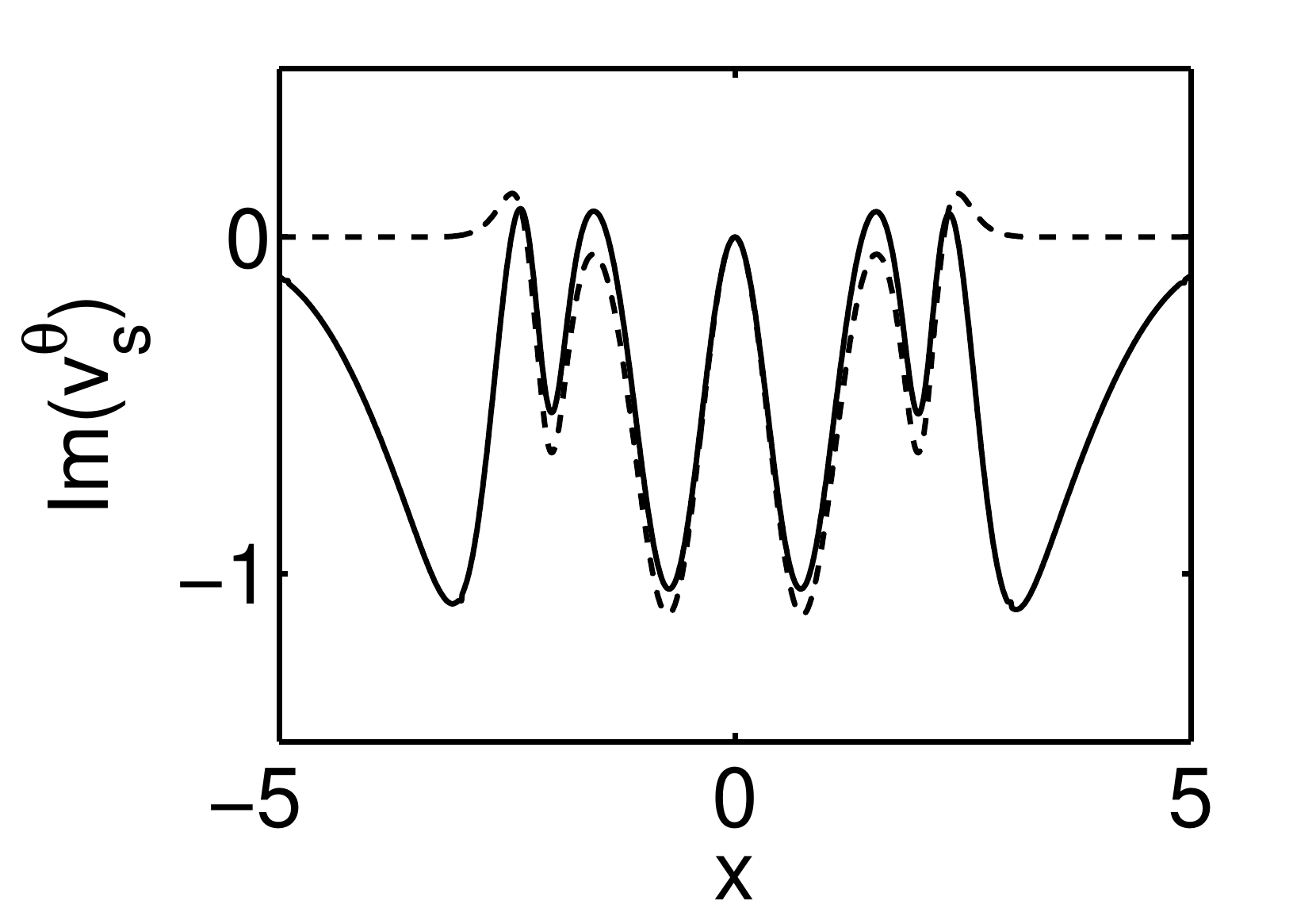}}
\caption{\label{fig:potandden} The real and imaginary part of the complex Kohn-Sham potential for the LER of 2 soft-Coulomb 
interacting electrons in the model potential, Eq.\ref{e:Hamiltonian}.  The dashed lines are the real and imaginary part of the complex-scaled parent 
potential $\tilde{v}(x)$. ($\theta = 0.35$, $a=4$, $b=0.5$, $c=4$, $d=2$, and $\lambda=1$).}
\end{figure}

{\underline {\em Exact KS solution}}:
Two non-interacting electrons in the 
potential indicated by solid lines in Fig.\ref{fig:potandden} have the same $n_\theta(x)$ as calculated above to one part in $10^6$ (in the sense that the space
integral of the square of the difference between their real or imaginary parts is less than $10^6$). When $n_\theta(x)$ is 
set to integrate to the number of electrons ($2$, here), we verify this potential is given by: 
\begin{equation}
 v_{s}^\theta(x) = e^{-2 i \theta} \frac{\nabla^2 \sqrt{n_\theta(x)}}{2 \sqrt{n_\theta(x)}} - \varepsilon_H+2i\tau_H^{-1}~~,
\label{eq:invert}
\end{equation}
where $\varepsilon_H-2i\tau_H^{-1}$ is the highest occupied complex orbital energy (in this case the only one), in exact analogy to real KS-potentials for bound 2-electron 
systems. 
Without an explicit expression for $E\xc^\theta[n_\theta]$, however, the total energy cannot be calculated via Eq.~\ref{e:hypothesis}.  Related work by Ernzerhof \cite{E06} and physical intuition suggest that bound ground-state functionals are applicable here. They 
are, in any case, the most natural candidates. 

{\underline{{\em Exchange}}}:
Borrowing knowledge from bound 2-electron DFT, Eqs.~\ref{e:KS} and~\ref{eqn:vs} were solved employing $E\x^\theta[n_\theta]=-\half E\H^\theta [n_\theta]=-\half e^{-i\theta}E\H[n_\theta]$. 
The complex KS equations can be solved self-consistently with an initial guess for $n_\theta$. Using the non-interacting complex density,
the SCF calculations converged in 4-5 iterations. The resulting complex energies are plotted in Fig.~\ref{fig:engs} along with the exact results. For comparison,
we also plot the results from perturbation theory to first order in $\lambda$. The two yield identical answers for the resonance
energies, and extremely close for the lifetimes for all $\lambda$ in the range $0<\lambda<1$. 
Thus, neglecting correlation, we find the average error is $\sim$ 14\% for the real part and $\sim$ 35\% for the imaginary part of the total energy.  
We also compare with standard scattering calculations using the close-coupling equations
under the bound state approximation~\cite{A98, taylor}. The resonance energy is predicted by this method with an error 
of 22\%, comparable to our DFRT exchange-only results.
  
As in standard KS-DFT, total energies are given here by:
\begin{eqnarray}
E_\theta[n_\theta] &=& \sum_{i=1}^N\left(\varepsilon_i-2i\tau_i^{-1}\right)+E\Hx^\theta[n_\theta] \nonumber \\
 & & -\int d\br v\Hx^\theta n_\theta(\br)
\end{eqnarray}
We point out that the $\theta$-independence of the energy is preserved by the SCF procedure (see Table~\ref{tab:par2}).
As the grid-size increases the dependence on $\theta$ becomes negligible.  This is important,
because within a SCF DFRT calculation one is always solving the 1-body complex KS-equations.  For these equations, one should 
be able to efficiently use a large enough basis set or a fine enough grid to extinguish most of the numerical $\theta$ dependence. Thus, this well-known
drawback of the complex-scaling technique \cite{moiseyev1,reinhardt,simon} is outdone by the benefit of never having to deal with $N$-particle wavefunctions, 
but just 1-body (complex) densities.


\begin{table}
\begin{tabular}{lclclcl}
\quad Grid &$\theta$ & \quad $\mbox{Re}(E)$ & \quad   $\mbox{Im}(E) $ \\
\hline
(N = 299) & 0.27 & \quad $4.99895$ & \quad $-0.0149586$ \\
& 0.35 & \quad $4.99933$ & \quad $-0.0144161$ \\
& 0.43 & \quad $4.99962$ & \quad $-0.0139792$ \\
\hline
(N = 1299) & 0.27 & \quad $5.00182$ & \quad $-0.0161045$ \\
& 0.35 & \quad $5.00198$ & \quad $-0.0159848$ \\
& 0.43 & \quad $5.00200$ & \quad $-0.0159513$ \\
\hline
\end{tabular}
\caption{Two-electron resonance energy values in the model Hamiltonian of Eq.~\ref{e:Hamiltonian} calculated via exchange-only DFRT.  As 
the grid spacing decreases numerical dependence on $\theta$ practically disappears.  ($a=4$, $b=0.5$, $c=4$, $d=2$ and $\lambda=1$)}
\label{tab:par2}
\end{table}

\begin{figure}
\scalebox{0.265}{\includegraphics{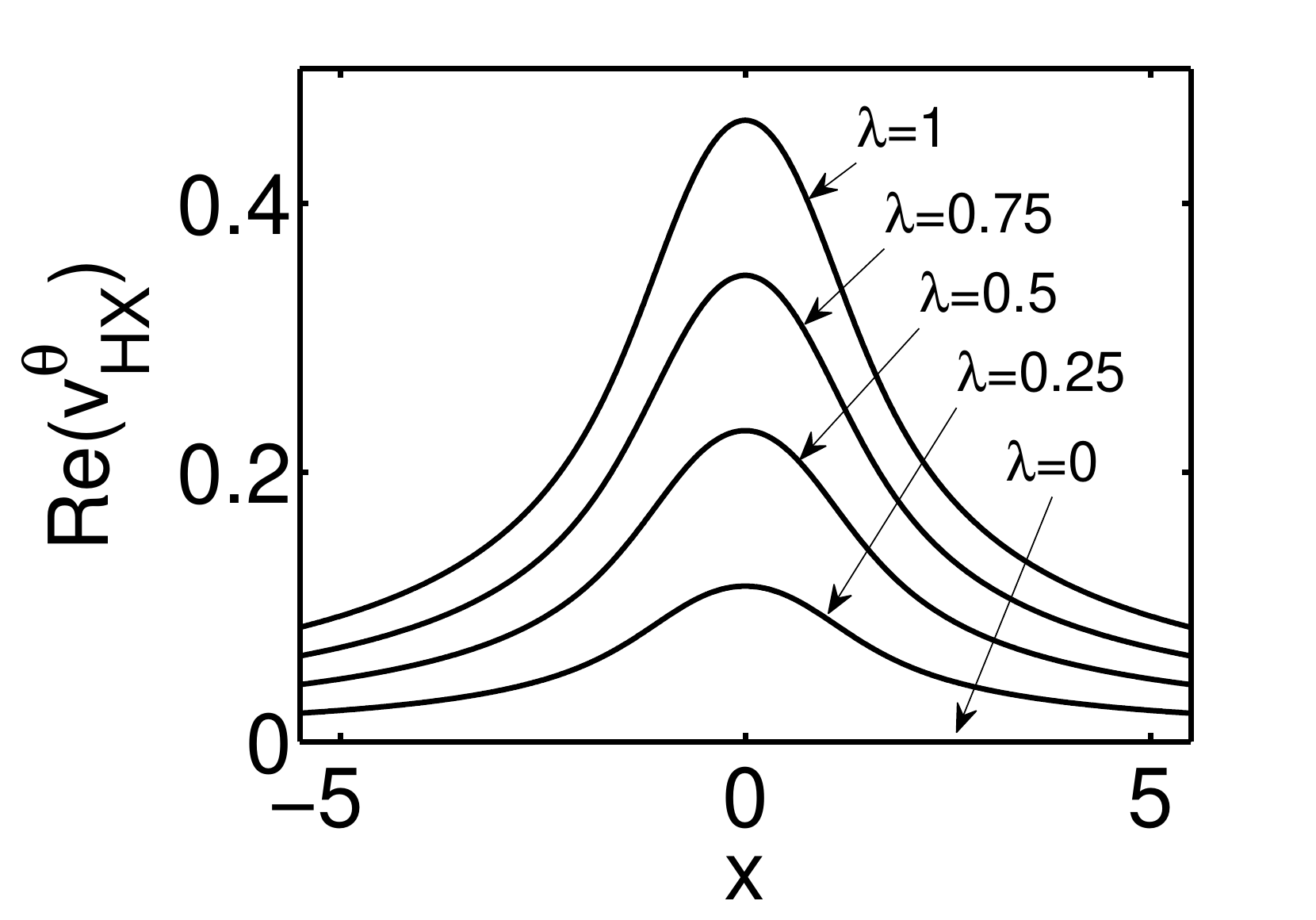}}
\hspace{-15pt}
\scalebox{0.265}{\includegraphics{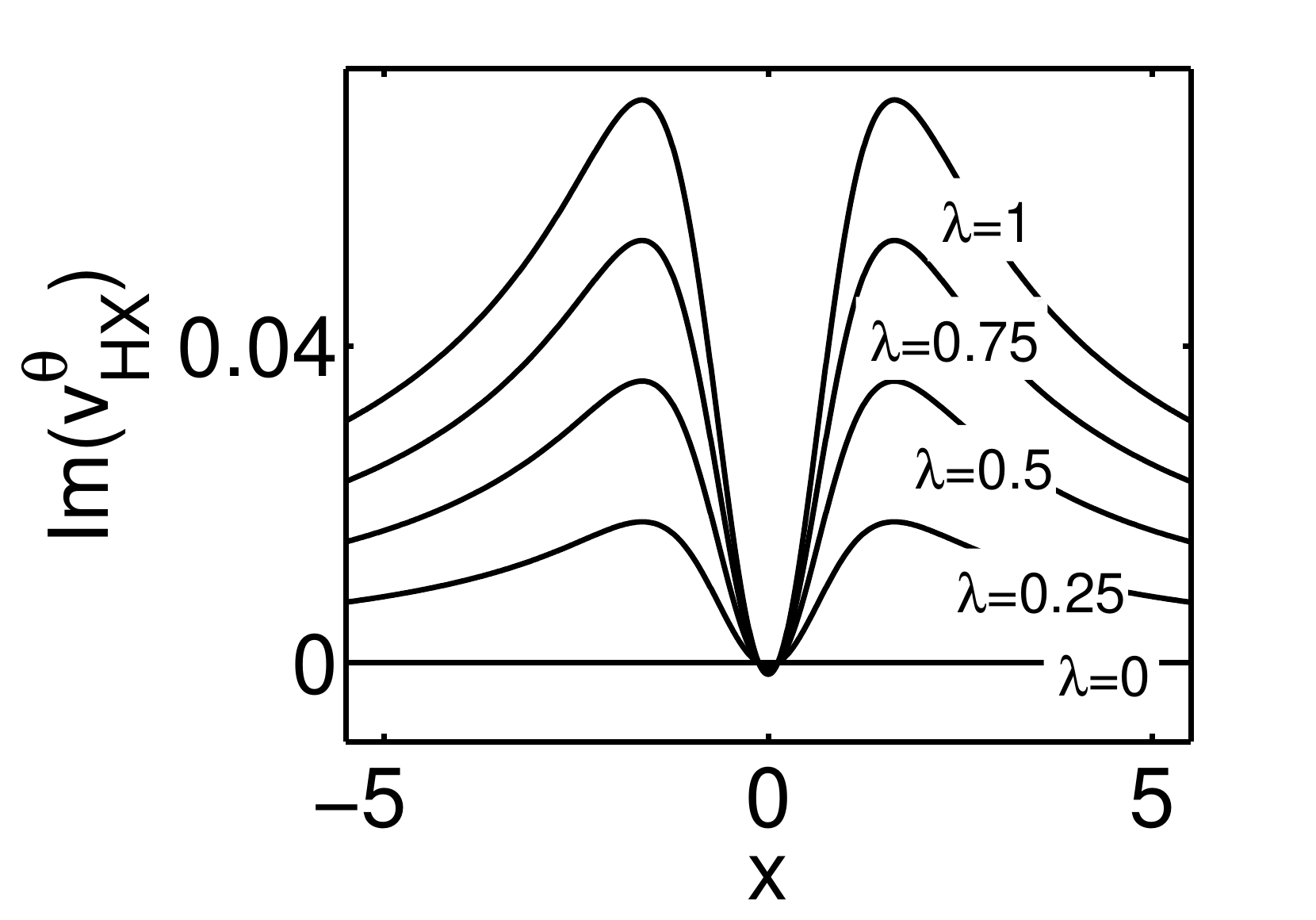}} 
\scalebox{0.265}{\includegraphics{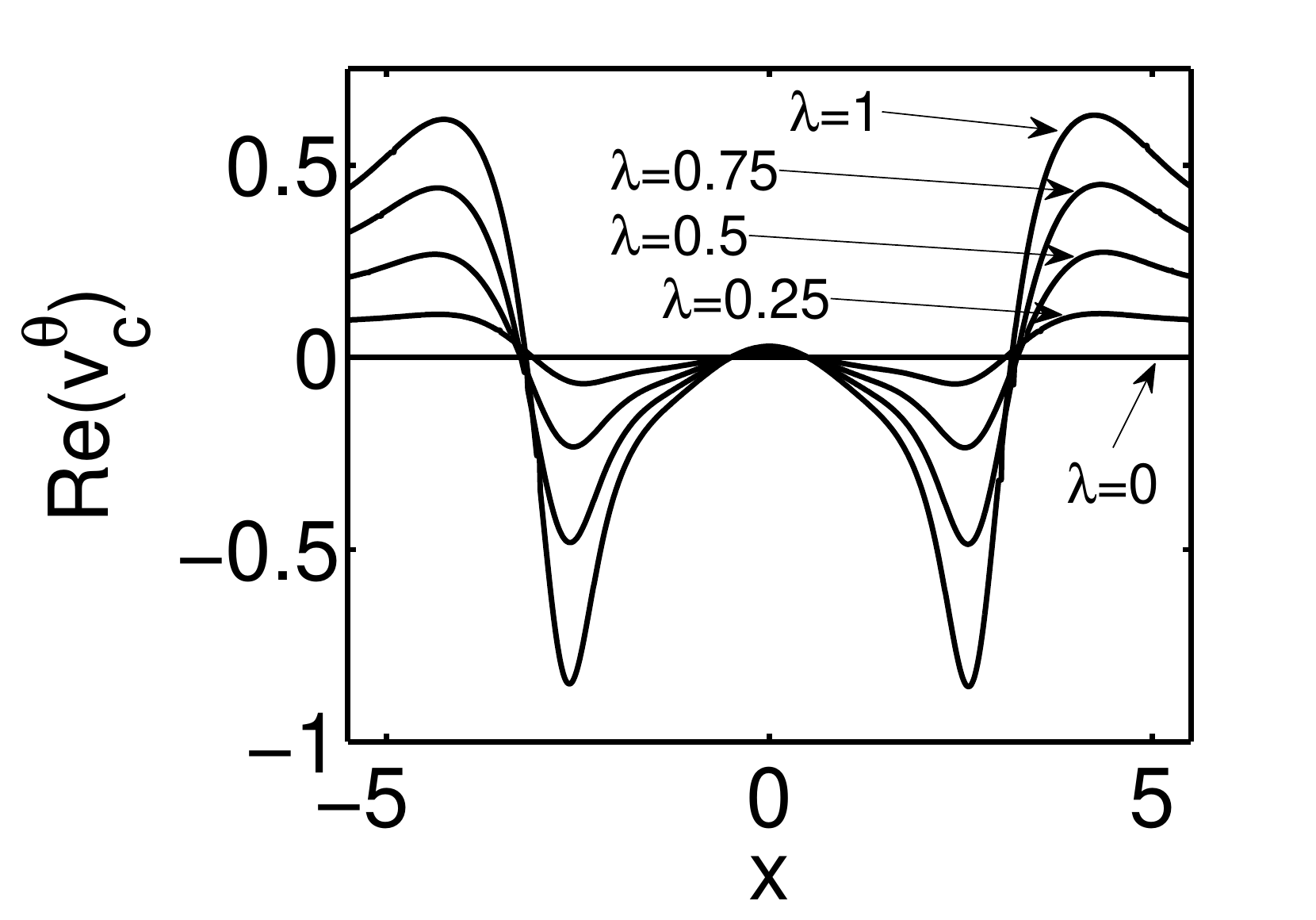}}
\hspace{-15pt}
\scalebox{0.265}{\includegraphics{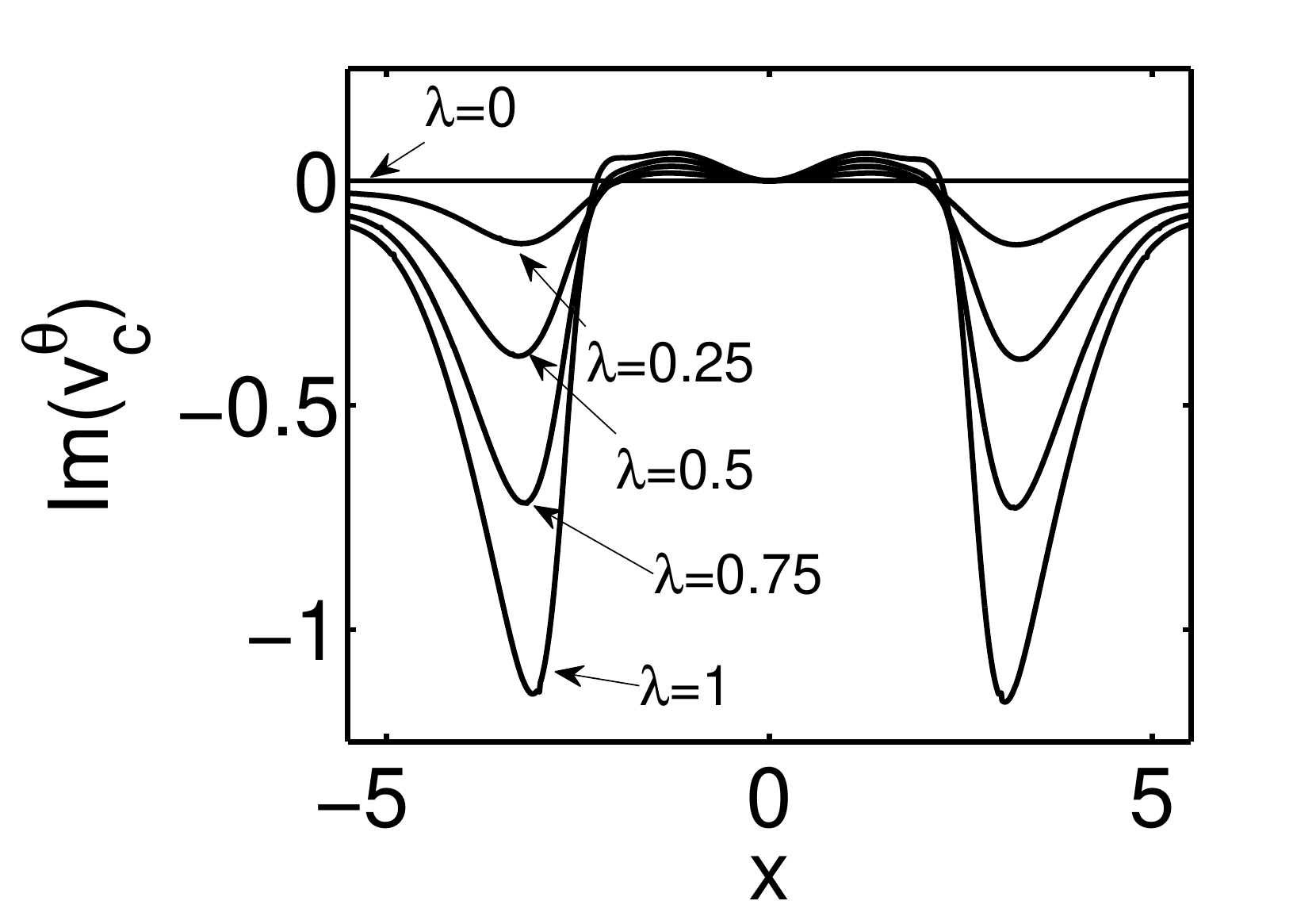}}
\caption{\label{fig:corr} The individual contributions to the Kohn-Sham potential from Hartree-exchange and correlation. ($\theta = 0.35$, $a=4$, $b=0.5$, $c=4$, and $d=2$)}
\end{figure}

{\underline{\em Correlation Potential}}:
It is of interest to calculate the exact correlation potential, which we do by subtracting the hartree-exchange contribution from the exact KS potential. The individual Hartree-exchange and correlation potentials are shown in Fig.~\ref{fig:corr}. To interpret the features in these complex potentials it is useful to distinguish between two regions. As the interaction between electrons is
turned on and $\lambda$ increases from 0 to 1, the region around the central well is shifted up in the real part of the Kohn-Sham potential.  This behavior is also seen in standard KS-DFT, and serves to shift up the position of the non-interacting orbital energies (in that case the real part of the orbital energies).  However, both the real and imaginary part of the complex Kohn-Sham potential have a second region outside the central well that shows a dramatic oscillatory structure arising purely from the fact that the state is unbound.  It is already known that the decaying oscillations in the tails of the complex LER wavefunction are governed by the lifetime of the resonance~\cite{PML90}.  These oscillations serve to produce the correct assymptotic behavior in the interacting complex density and thereby give the correct interacting lifetime when this density is used in the functional.  

The analog of Koopmans' theorem does not hold in DFRT. Although the ionization energy of our 2-electron system is strictly zero, it is tempting to define $I_\theta \equiv E_\theta(N=1)-E_\theta(N=2)$ and check whether it equals the highest occupied KS orbital energy. For the parameters used in Figs.1-4, $E_\theta(N=1)=1.629-0.003i$, $E_\theta(N=2)=4.127-0.014i$, but the exact KS eigenvalue is $2.065-0.006i$.  Clearly, DFRT provides an unambiguous prescription for the calculation of negative electron affinities. 

We are working on the implementation of DFRT to calculate the lifetime of molecular metastable anions. The method would also
be applicable to molecules connected to metallic leads, as in molecular electronics.  Ernzerhof and co-workers have developed an approach for that purpose where complex absorbing potentials are added within a complex-DFT framework \cite{E06, GEZ07}. However, we emphasize that the complex potentials in DFRT are the result of a variational calculation, and they are obtained self-consistently for the $N$-electron system treated as isolated, rather than added to the Hamiltonian from the start to model an open system. 

In addition, DFRT should be applicable to study shape and Feshbach resonances in low-energy electron scattering
processes \cite{palmer,S08,*S11,FGMS07} of growing interest in biological systems \cite{PCHS03,PS05,MBBF10}, atmospheric sciences, lasers, and astrophysics \cite{massey,LSBL02,S10,OSAW11}. 


In summary, DFRT provides an unambiguous prescription for calculating negative electron affinities based on a complex-scaled version of standard ground state DFT.  This complex-scaled version has been cast in a way that is analogous in practice to KS-DFT. Results on a model system suggest that the same machinery that has been developed for KS-DFT yields accurate resonance energies and lifetimes in DFRT. It remains to be seen if common approximations to $E_{XC}[n]$ are able to capture the important effects that determine properties of real transient anions. A more detailed study of the complex density function and various DFRT identities is forthcoming.

\begin{acknowledgments}
Acknowledgment is made to the Donors of the American Chemical Society Petroleum Research Fund for support of this
research under grant No.PRF\# 49599-DNI6.
\end{acknowledgments}

\bibliography{ksres2}

\begin{thebibliography}{35}%
\makeatletter
\providecommand \@ifxundefined [1]{%
 \@ifx{#1\undefined}
}%
\providecommand \@ifnum [1]{%
 \ifnum #1\expandafter \@firstoftwo
 \else \expandafter \@secondoftwo
 \fi
}%
\providecommand \@ifx [1]{%
 \ifx #1\expandafter \@firstoftwo
 \else \expandafter \@secondoftwo
 \fi
}%
\providecommand \natexlab [1]{#1}%
\providecommand \enquote  [1]{``#1''}%
\providecommand \bibnamefont  [1]{#1}%
\providecommand \bibfnamefont [1]{#1}%
\providecommand \citenamefont [1]{#1}%
\providecommand \href@noop [0]{\@secondoftwo}%
\providecommand \href [0]{\begingroup \@sanitize@url \@href}%
\providecommand \@href[1]{\@@startlink{#1}\@@href}%
\providecommand \@@href[1]{\endgroup#1\@@endlink}%
\providecommand \@sanitize@url [0]{\catcode `\\12\catcode `\$12\catcode
  `\&12\catcode `\#12\catcode `\^12\catcode `\_12\catcode `\%12\relax}%
\providecommand \@@startlink[1]{}%
\providecommand \@@endlink[0]{}%
\providecommand \url  [0]{\begingroup\@sanitize@url \@url }%
\providecommand \@url [1]{\endgroup\@href {#1}{\urlprefix }}%
\providecommand \urlprefix  [0]{URL }%
\providecommand \Eprint [0]{\href }%
\@ifxundefined \urlstyle {%
  \providecommand \doi  [0]{\begingroup \@sanitize@url \@doi}%
  \providecommand \@doi [1]{\endgroup \@@startlink {\doibase
  #1}doi:\discretionary {}{}{}#1\@@endlink }%
}{%
  \providecommand \doi  [0]{doi:\discretionary{}{}{}\begingroup
  \urlstyle{rm}\Url }%
}%
\providecommand \doibase [0]{http://dx.doi.org/}%
\providecommand \Doi [0]{\begingroup \@sanitize@url \@Doi }%
\providecommand \@Doi  [1]{\endgroup\@@startlink{\doibase#1}\@@Doi}%
\providecommand \@@Doi [1]{#1\@@endlink}%
\providecommand \selectlanguage [0]{\@gobble}%
\providecommand \bibinfo  [0]{\@secondoftwo}%
\providecommand \bibfield  [0]{\@secondoftwo}%
\providecommand \translation [1]{[#1]}%
\providecommand \BibitemOpen [0]{}%
\providecommand \bibitemStop [0]{}%
\providecommand \bibitemNoStop [0]{.\EOS\space}%
\providecommand \EOS [0]{\spacefactor3000\relax}%
\providecommand \BibitemShut  [1]{\csname bibitem#1\endcsname}%
\bibitem [{\citenamefont {Hohenberg}\ and\ \citenamefont
  {Kohn}(1964)}]{hohenberg}%
  \BibitemOpen
  \bibfield  {author} {\bibinfo {author} {\bibfnamefont {P.}~\bibnamefont
  {Hohenberg}}\ and\ \bibinfo {author} {\bibfnamefont {W.}~\bibnamefont
  {Kohn}},\ }\href@noop {} {\bibfield  {journal} {\bibinfo  {journal} {Phys.
  Rev.},\ }\textbf {\bibinfo {volume} {136}},\ \bibinfo {pages} {B864}
  (\bibinfo {year} {1964})}\BibitemShut {NoStop}%
\bibitem [{\citenamefont {Kohn}\ and\ \citenamefont {Sham}(1965)}]{kohn}%
  \BibitemOpen
  \bibfield  {author} {\bibinfo {author} {\bibfnamefont {W.}~\bibnamefont
  {Kohn}}\ and\ \bibinfo {author} {\bibfnamefont {L.~J.}\ \bibnamefont
  {Sham}},\ }\href@noop {} {\bibfield  {journal} {\bibinfo  {journal} {Phys.
  Rev.},\ }\textbf {\bibinfo {volume} {140}},\ \bibinfo {pages} {A1133}
  (\bibinfo {year} {1965})}\BibitemShut {NoStop}%
\bibitem [{\citenamefont {Parr}\ and\ \citenamefont {Yang}(1994)}]{parr}%
  \BibitemOpen
  \bibfield  {author} {\bibinfo {author} {\bibfnamefont {R.~G.}\ \bibnamefont
  {Parr}}\ and\ \bibinfo {author} {\bibfnamefont {W.}~\bibnamefont {Yang}},\
  }\href@noop {} {\emph {\bibinfo {title} {Density-Functional Theory of Atoms
  and Molecules}}}\ (\bibinfo  {publisher} {Oxford University Press},\ \bibinfo
  {address} {Oxford},\ \bibinfo {year} {1994})\BibitemShut {NoStop}%
\bibitem [{\citenamefont {Martin}(2004)}]{Martin}%
  \BibitemOpen
  \bibfield  {author} {\bibinfo {author} {\bibfnamefont {R.~M.}\ \bibnamefont
  {Martin}},\ }\href@noop {} {\emph {\bibinfo {title} {Electronic Structure:
  Basic Theory and Practical Methods}}}\ (\bibinfo  {publisher} {Cambridge
  University Press},\ \bibinfo {address} {London},\ \bibinfo {year}
  {2004})\BibitemShut {NoStop}%
\bibitem [{\citenamefont {Runge}\ and\ \citenamefont {Gross}(1984)}]{RG84}%
  \BibitemOpen
  \bibfield  {author} {\bibinfo {author} {\bibfnamefont {E.}~\bibnamefont
  {Runge}}\ and\ \bibinfo {author} {\bibfnamefont {E.~K.~U.}\ \bibnamefont
  {Gross}},\ }\href@noop {} {\bibfield  {journal} {\bibinfo  {journal} {Phys.
  Rev. Lett.},\ }\textbf {\bibinfo {volume} {52}},\ \bibinfo {pages} {997}
  (\bibinfo {year} {1984})}\BibitemShut {NoStop}%
\bibitem [{\citenamefont {Marques}\ \emph {et~al.}(2006)\citenamefont
  {Marques}, \citenamefont {Ullrich}, \citenamefont {Nogueira}, \citenamefont
  {Rubio}, \citenamefont {Burke},\ and\ \citenamefont {Gross}}]{TDDFT}%
  \BibitemOpen
  \bibfield  {author} {\bibinfo {author} {\bibfnamefont {M.~A.~L.}\
  \bibnamefont {Marques}}, \bibinfo {author} {\bibfnamefont {C.~A.}\
  \bibnamefont {Ullrich}}, \bibinfo {author} {\bibfnamefont {F.}~\bibnamefont
  {Nogueira}}, \bibinfo {author} {\bibfnamefont {A.}~\bibnamefont {Rubio}},
  \bibinfo {author} {\bibfnamefont {K.}~\bibnamefont {Burke}}, \ and\ \bibinfo
  {author} {\bibfnamefont {E.~K.~U.}\ \bibnamefont {Gross}},\ }\href@noop {}
  {\emph {\bibinfo {title} {Time-Dependent Density Functional Theory}}}\
  (\bibinfo  {publisher} {Springer},\ \bibinfo {address} {Verlag Berlin
  Heidelberg},\ \bibinfo {year} {2006})\BibitemShut {NoStop}%
\bibitem [{\citenamefont {Lee}\ \emph {et~al.}(2010)\citenamefont {Lee},
  \citenamefont {Furche},\ and\ \citenamefont {Burke}}]{LFB10}%
  \BibitemOpen
  \bibfield  {author} {\bibinfo {author} {\bibfnamefont {D.}~\bibnamefont
  {Lee}}, \bibinfo {author} {\bibfnamefont {F.}~\bibnamefont {Furche}}, \ and\
  \bibinfo {author} {\bibfnamefont {K.}~\bibnamefont {Burke}},\ }\href@noop {}
  {\bibfield  {journal} {\bibinfo  {journal} {J. Phys. Chem. Lett.},\ }\textbf
  {\bibinfo {volume} {1}},\ \bibinfo {pages} {2124} (\bibinfo {year}
  {2010})}\BibitemShut {NoStop}%
\bibitem [{\citenamefont {Kim}\ \emph {et~al.}(2011)\citenamefont {Kim},
  \citenamefont {Sim},\ and\ \citenamefont {Burke}}]{KSB11}%
  \BibitemOpen
  \bibfield  {author} {\bibinfo {author} {\bibfnamefont {M.}~\bibnamefont
  {Kim}}, \bibinfo {author} {\bibfnamefont {E.}~\bibnamefont {Sim}}, \ and\
  \bibinfo {author} {\bibfnamefont {K.}~\bibnamefont {Burke}},\ }\href@noop {}
  {\bibfield  {journal} {\bibinfo  {journal} {J. Chem. Phys.},\ }\textbf
  {\bibinfo {volume} {134}},\ \bibinfo {pages} {171103} (\bibinfo {year}
  {2011})}\BibitemShut {NoStop}%
\bibitem [{\citenamefont {Krueger}\ and\ \citenamefont {Maitra}(2009)}]{KM09}%
  \BibitemOpen
  \bibfield  {author} {\bibinfo {author} {\bibfnamefont {A.~J.}\ \bibnamefont
  {Krueger}}\ and\ \bibinfo {author} {\bibfnamefont {N.~T.}\ \bibnamefont
  {Maitra}},\ }\href@noop {} {\bibfield  {journal} {\bibinfo  {journal} {Phys.
  Chem. Chem. Phys.},\ }\textbf {\bibinfo {volume} {11}},\ \bibinfo {pages}
  {4655} (\bibinfo {year} {2009})}\BibitemShut {NoStop}%
\bibitem [{\citenamefont {van Faassen}\ and\ \citenamefont
  {Burke}(2006){\natexlab{a}}}]{FB06}%
  \BibitemOpen
  \bibfield  {author} {\bibinfo {author} {\bibfnamefont {M.}~\bibnamefont {van
  Faassen}}\ and\ \bibinfo {author} {\bibfnamefont {K.}~\bibnamefont {Burke}},\
  }\href@noop {} {\bibfield  {journal} {\bibinfo  {journal} {J. Chem. Phys.},\
  }\textbf {\bibinfo {volume} {124}},\ \bibinfo {pages} {094102} (\bibinfo
  {year} {2006}{\natexlab{a}})}\BibitemShut {NoStop}%
\bibitem [{\citenamefont {van Faassen}\ and\ \citenamefont
  {Burke}(2006){\natexlab{b}}}]{FB06b}%
  \BibitemOpen
  \bibfield  {author} {\bibinfo {author} {\bibfnamefont {M.}~\bibnamefont {van
  Faassen}}\ and\ \bibinfo {author} {\bibfnamefont {K.}~\bibnamefont {Burke}},\
  }\href@noop {} {\bibfield  {journal} {\bibinfo  {journal} {Chem. Phys.
  Lett.},\ }\textbf {\bibinfo {volume} {431}},\ \bibinfo {pages} {410}
  (\bibinfo {year} {2006}{\natexlab{b}})}\BibitemShut {NoStop}%
\bibitem [{\citenamefont {van Faassen}\ and\ \citenamefont
  {Burke}(2009)}]{FB09}%
  \BibitemOpen
  \bibfield  {author} {\bibinfo {author} {\bibfnamefont {M.}~\bibnamefont {van
  Faassen}}\ and\ \bibinfo {author} {\bibfnamefont {K.}~\bibnamefont {Burke}},\
  }\href@noop {} {\bibfield  {journal} {\bibinfo  {journal} {Phys. Chem. Chem.
  Phys.},\ }\textbf {\bibinfo {volume} {11}},\ \bibinfo {pages} {4437}
  (\bibinfo {year} {2009})}\BibitemShut {NoStop}%
\bibitem [{\citenamefont {Whitenack}\ and\ \citenamefont
  {Wasserman}(2010)}]{WW10}%
  \BibitemOpen
  \bibfield  {author} {\bibinfo {author} {\bibfnamefont {D.~L.}\ \bibnamefont
  {Whitenack}}\ and\ \bibinfo {author} {\bibfnamefont {A.}~\bibnamefont
  {Wasserman}},\ }\href@noop {} {\bibfield  {journal} {\bibinfo  {journal} {J.
  Phys. Chem. Lett.},\ }\textbf {\bibinfo {volume} {1}},\ \bibinfo {pages}
  {407} (\bibinfo {year} {2010})}\BibitemShut {NoStop}%
\bibitem [{\citenamefont {Moiseyev}(2011)}]{M11}%
  \BibitemOpen
  \bibfield  {author} {\bibinfo {author} {\bibfnamefont {N.}~\bibnamefont
  {Moiseyev}},\ }\href@noop {} {\emph {\bibinfo {title} {Non-Hermitian Quantum
  Mechanics}}}\ (\bibinfo  {publisher} {Cambridge University Press},\ \bibinfo
  {address} {London},\ \bibinfo {year} {2011})\BibitemShut {NoStop}%
\bibitem [{\citenamefont {Ernzerhof}(2006)}]{E06}%
  \BibitemOpen
  \bibfield  {author} {\bibinfo {author} {\bibfnamefont {M.}~\bibnamefont
  {Ernzerhof}},\ }\href@noop {} {\bibfield  {journal} {\bibinfo  {journal} {J.
  Chem. Phys.},\ }\textbf {\bibinfo {volume} {125}},\ \bibinfo {pages} {124104}
  (\bibinfo {year} {2006})}\BibitemShut {NoStop}%
\bibitem [{\citenamefont {Wasserman}\ and\ \citenamefont
  {Moiseyev}(2007)}]{WM07}%
  \BibitemOpen
  \bibfield  {author} {\bibinfo {author} {\bibfnamefont {A.}~\bibnamefont
  {Wasserman}}\ and\ \bibinfo {author} {\bibfnamefont {N.}~\bibnamefont
  {Moiseyev}},\ }\href@noop {} {\bibfield  {journal} {\bibinfo  {journal}
  {Phys. Rev. Lett.},\ }\textbf {\bibinfo {volume} {98}},\ \bibinfo {pages}
  {093003} (\bibinfo {year} {2007})}\BibitemShut {NoStop}%
\bibitem [{\citenamefont {Chu}(1990)}]{chu}%
  \BibitemOpen
  \bibfield  {author} {\bibinfo {author} {\bibfnamefont {S.}~\bibnamefont
  {Chu}},\ }\href@noop {} {\bibfield  {journal} {\bibinfo  {journal} {Chem.
  Phys. Lett.},\ }\textbf {\bibinfo {volume} {167}},\ \bibinfo {pages} {155}
  (\bibinfo {year} {1990})}\BibitemShut {NoStop}%
\bibitem [{\citenamefont {Adhikari}(1998)}]{A98}%
  \BibitemOpen
  \bibfield  {author} {\bibinfo {author} {\bibfnamefont {S.~K.}\ \bibnamefont
  {Adhikari}},\ }\href@noop {} {\emph {\bibinfo {title} {Variational Principles
  and the Numerical Solution of Scattering Problems}}}\ (\bibinfo  {publisher}
  {John Wiley and Sons, Inc.},\ \bibinfo {address} {New Jersey},\ \bibinfo
  {year} {1998})\BibitemShut {NoStop}%
\bibitem [{\citenamefont {Taylor}(1972)}]{taylor}%
  \BibitemOpen
  \bibfield  {author} {\bibinfo {author} {\bibfnamefont {J.~R.}\ \bibnamefont
  {Taylor}},\ }\href@noop {} {\emph {\bibinfo {title} {Scattering Theory}}}\
  (\bibinfo  {publisher} {Dover Publications, Inc.},\ \bibinfo {address} {New
  York},\ \bibinfo {year} {1972})\BibitemShut {NoStop}%
\bibitem [{\citenamefont {Moiseyev}\ \emph {et~al.}(1978)\citenamefont
  {Moiseyev}, \citenamefont {Certain},\ and\ \citenamefont
  {Weinhold}}]{moiseyev1}%
  \BibitemOpen
  \bibfield  {author} {\bibinfo {author} {\bibfnamefont {N.}~\bibnamefont
  {Moiseyev}}, \bibinfo {author} {\bibfnamefont {P.~R.}\ \bibnamefont
  {Certain}}, \ and\ \bibinfo {author} {\bibfnamefont {F.}~\bibnamefont
  {Weinhold}},\ }\href@noop {} {\bibfield  {journal} {\bibinfo  {journal} {Mol.
  Phys.},\ }\textbf {\bibinfo {volume} {36}},\ \bibinfo {pages} {1613}
  (\bibinfo {year} {1978})}\BibitemShut {NoStop}%
\bibitem [{\citenamefont {Reinhardt}(1982)}]{reinhardt}%
  \BibitemOpen
  \bibfield  {author} {\bibinfo {author} {\bibfnamefont {W.~P.}\ \bibnamefont
  {Reinhardt}},\ }\href@noop {} {\bibfield  {journal} {\bibinfo  {journal}
  {Ann. Rev. Phys. Chem.},\ }\textbf {\bibinfo {volume} {33}},\ \bibinfo
  {pages} {223} (\bibinfo {year} {1982})}\BibitemShut {NoStop}%
\bibitem [{\citenamefont {Simon}(1973)}]{simon}%
  \BibitemOpen
  \bibfield  {author} {\bibinfo {author} {\bibfnamefont {B.}~\bibnamefont
  {Simon}},\ }\href@noop {} {\bibfield  {journal} {\bibinfo  {journal} {Ann.
  Math.},\ }\textbf {\bibinfo {volume} {97}},\ \bibinfo {pages} {247} (\bibinfo
  {year} {1973})}\BibitemShut {NoStop}%
\bibitem [{\citenamefont {Peskin}\ \emph {et~al.}(1990)\citenamefont {Peskin},
  \citenamefont {Moiseyev},\ and\ \citenamefont {Lefebvre}}]{PML90}%
  \BibitemOpen
  \bibfield  {author} {\bibinfo {author} {\bibfnamefont {U.}~\bibnamefont
  {Peskin}}, \bibinfo {author} {\bibfnamefont {N.}~\bibnamefont {Moiseyev}}, \
  and\ \bibinfo {author} {\bibfnamefont {R.}~\bibnamefont {Lefebvre}},\
  }\href@noop {} {\bibfield  {journal} {\bibinfo  {journal} {J. Chem. Phys.},\
  }\textbf {\bibinfo {volume} {92}},\ \bibinfo {pages} {2902} (\bibinfo {year}
  {1990})}\BibitemShut {NoStop}%
\bibitem [{\citenamefont {Goyer}\ \emph {et~al.}(2007)\citenamefont {Goyer},
  \citenamefont {Ernzerhof},\ and\ \citenamefont {Zhuang}}]{GEZ07}%
  \BibitemOpen
  \bibfield  {author} {\bibinfo {author} {\bibfnamefont {F.}~\bibnamefont
  {Goyer}}, \bibinfo {author} {\bibfnamefont {M.}~\bibnamefont {Ernzerhof}}, \
  and\ \bibinfo {author} {\bibfnamefont {M.}~\bibnamefont {Zhuang}},\
  }\href@noop {} {\bibfield  {journal} {\bibinfo  {journal} {J. Chem. Phys.},\
  }\textbf {\bibinfo {volume} {126}},\ \bibinfo {pages} {144104} (\bibinfo
  {year} {2007})}\BibitemShut {NoStop}%
\bibitem [{\citenamefont {Palmer}\ and\ \citenamefont {Rous}(1992)}]{palmer}%
  \BibitemOpen
  \bibfield  {author} {\bibinfo {author} {\bibfnamefont {R.~E.}\ \bibnamefont
  {Palmer}}\ and\ \bibinfo {author} {\bibfnamefont {P.~J.}\ \bibnamefont
  {Rous}},\ }\href@noop {} {\bibfield  {journal} {\bibinfo  {journal} {Rev.
  Mod. Phys.},\ }\textbf {\bibinfo {volume} {64}},\ \bibinfo {pages} {383}
  (\bibinfo {year} {1992})}\BibitemShut {NoStop}%
\bibitem [{\citenamefont {Simons}(2008)}]{S08}%
  \BibitemOpen
  \bibfield  {author} {\bibinfo {author} {\bibfnamefont {J.}~\bibnamefont
  {Simons}},\ }\href@noop {} {\bibfield  {journal} {\bibinfo  {journal} {J.
  Phys. Chem. A},\ }\textbf {\bibinfo {volume} {112}},\ \bibinfo {pages} {6401}
  (\bibinfo {year} {2008})}\BibitemShut {NoStop}%
\bibitem [{\citenamefont {Simons}(2011)}]{S11}%
  \BibitemOpen
  \bibfield  {author} {\bibinfo {author} {\bibfnamefont {J.}~\bibnamefont
  {Simons}},\ }\href@noop {} {\bibfield  {journal} {\bibinfo  {journal} {Annu.
  Rev. Phys. Chem.},\ }\textbf {\bibinfo {volume} {67}},\ \bibinfo {pages}
  {107} (\bibinfo {year} {2011})}\BibitemShut {NoStop}%
\bibitem [{\citenamefont {Field}\ \emph {et~al.}(2007)\citenamefont {Field},
  \citenamefont {Graupner}, \citenamefont {Mauracher}, \citenamefont {Scheier},
  \citenamefont {Bacher}, \citenamefont {Denifl}, \citenamefont {Zappa},\ and\
  \citenamefont {Mark}}]{FGMS07}%
  \BibitemOpen
  \bibfield  {author} {\bibinfo {author} {\bibfnamefont {T.~A.}\ \bibnamefont
  {Field}}, \bibinfo {author} {\bibfnamefont {K.}~\bibnamefont {Graupner}},
  \bibinfo {author} {\bibfnamefont {A.}~\bibnamefont {Mauracher}}, \bibinfo
  {author} {\bibfnamefont {P.}~\bibnamefont {Scheier}}, \bibinfo {author}
  {\bibfnamefont {A.}~\bibnamefont {Bacher}}, \bibinfo {author} {\bibfnamefont
  {S.}~\bibnamefont {Denifl}}, \bibinfo {author} {\bibfnamefont
  {F.}~\bibnamefont {Zappa}}, \ and\ \bibinfo {author} {\bibfnamefont {T.~D.}\
  \bibnamefont {Mark}},\ }\href@noop {} {\bibfield  {journal} {\bibinfo
  {journal} {J. Phys.: Conf. Ser.},\ }\textbf {\bibinfo {volume} {88}},\
  \bibinfo {pages} {012029} (\bibinfo {year} {2007})}\BibitemShut {NoStop}%
\bibitem [{\citenamefont {Pan}\ \emph {et~al.}(2003)\citenamefont {Pan},
  \citenamefont {Cloutier}, \citenamefont {Hunting},\ and\ \citenamefont
  {Sanche}}]{PCHS03}%
  \BibitemOpen
  \bibfield  {author} {\bibinfo {author} {\bibfnamefont {X.}~\bibnamefont
  {Pan}}, \bibinfo {author} {\bibfnamefont {P.}~\bibnamefont {Cloutier}},
  \bibinfo {author} {\bibfnamefont {D.}~\bibnamefont {Hunting}}, \ and\
  \bibinfo {author} {\bibfnamefont {L.}~\bibnamefont {Sanche}},\ }\href@noop {}
  {\bibfield  {journal} {\bibinfo  {journal} {Phys. Rev. Lett.},\ }\textbf
  {\bibinfo {volume} {90}},\ \bibinfo {pages} {208102} (\bibinfo {year}
  {2003})}\BibitemShut {NoStop}%
\bibitem [{\citenamefont {Pan}\ and\ \citenamefont {Sanche}(2005)}]{PS05}%
  \BibitemOpen
  \bibfield  {author} {\bibinfo {author} {\bibfnamefont {X.}~\bibnamefont
  {Pan}}\ and\ \bibinfo {author} {\bibfnamefont {L.}~\bibnamefont {Sanche}},\
  }\href@noop {} {\bibfield  {journal} {\bibinfo  {journal} {Phys. Rev.
  Lett.},\ }\textbf {\bibinfo {volume} {94}},\ \bibinfo {pages} {198104}
  (\bibinfo {year} {2005})}\BibitemShut {NoStop}%
\bibitem [{\citenamefont {Mucke}\ \emph {et~al.}(2010)\citenamefont {Mucke},
  \citenamefont {Braune}, \citenamefont {Barth}, \citenamefont {Frstel},
  \citenamefont {Lischke}, \citenamefont {Ulrich}, \citenamefont {Arion},
  \citenamefont {Becker}, \citenamefont {Bradshaw},\ and\ \citenamefont
  {Hergenhahn}}]{MBBF10}%
  \BibitemOpen
  \bibfield  {author} {\bibinfo {author} {\bibfnamefont {M.}~\bibnamefont
  {Mucke}}, \bibinfo {author} {\bibfnamefont {M.}~\bibnamefont {Braune}},
  \bibinfo {author} {\bibfnamefont {S.}~\bibnamefont {Barth}}, \bibinfo
  {author} {\bibfnamefont {M.}~\bibnamefont {Frstel}}, \bibinfo {author}
  {\bibfnamefont {T.}~\bibnamefont {Lischke}}, \bibinfo {author} {\bibfnamefont
  {V.}~\bibnamefont {Ulrich}}, \bibinfo {author} {\bibfnamefont
  {T.}~\bibnamefont {Arion}}, \bibinfo {author} {\bibfnamefont
  {U.}~\bibnamefont {Becker}}, \bibinfo {author} {\bibfnamefont
  {A.}~\bibnamefont {Bradshaw}}, \ and\ \bibinfo {author} {\bibfnamefont
  {U.}~\bibnamefont {Hergenhahn}},\ }\href@noop {} {\bibfield  {journal}
  {\bibinfo  {journal} {Nature Physics},\ }\textbf {\bibinfo {volume} {6}},\
  \bibinfo {pages} {143} (\bibinfo {year} {2010})}\BibitemShut {NoStop}%
\bibitem [{\citenamefont {Massey}(1950)}]{massey}%
  \BibitemOpen
  \bibfield  {author} {\bibinfo {author} {\bibfnamefont {H.~S.~W.}\
  \bibnamefont {Massey}},\ }\href@noop {} {\emph {\bibinfo {title} {Negative
  Ions}}}\ (\bibinfo  {publisher} {Cambridge University Press},\ \bibinfo
  {address} {London},\ \bibinfo {year} {1950})\BibitemShut {NoStop}%
\bibitem [{\citenamefont {Liang1}\ \emph {et~al.}(2002)\citenamefont {Liang1},
  \citenamefont {Shores}, \citenamefont {Bockrath}, \citenamefont {Long},\ and\
  \citenamefont {Park}}]{LSBL02}%
  \BibitemOpen
  \bibfield  {author} {\bibinfo {author} {\bibfnamefont {W.}~\bibnamefont
  {Liang1}}, \bibinfo {author} {\bibfnamefont {M.~P.}\ \bibnamefont {Shores}},
  \bibinfo {author} {\bibfnamefont {M.}~\bibnamefont {Bockrath}}, \bibinfo
  {author} {\bibfnamefont {J.~R.}\ \bibnamefont {Long}}, \ and\ \bibinfo
  {author} {\bibfnamefont {H.}~\bibnamefont {Park}},\ }\href@noop {} {\bibfield
   {journal} {\bibinfo  {journal} {Nature},\ }\textbf {\bibinfo {volume}
  {417}},\ \bibinfo {pages} {725} (\bibinfo {year} {2002})}\BibitemShut
  {NoStop}%
\bibitem [{\citenamefont {Strelkov}(2010)}]{S10}%
  \BibitemOpen
  \bibfield  {author} {\bibinfo {author} {\bibfnamefont {V.}~\bibnamefont
  {Strelkov}},\ }\href@noop {} {\bibfield  {journal} {\bibinfo  {journal}
  {Phys. Rev. Lett.},\ }\textbf {\bibinfo {volume} {104}},\ \bibinfo {pages}
  {123901} (\bibinfo {year} {2010})}\BibitemShut {NoStop}%
\bibitem [{\citenamefont {Olivares-Amaya}\ \emph {et~al.}(2011)\citenamefont
  {Olivares-Amaya}, \citenamefont {Stopa}, \citenamefont {Andrade},
  \citenamefont {Watson},\ and\ \citenamefont {Aspuru-Guzik}}]{OSAW11}%
  \BibitemOpen
  \bibfield  {author} {\bibinfo {author} {\bibfnamefont {R.}~\bibnamefont
  {Olivares-Amaya}}, \bibinfo {author} {\bibfnamefont {M.}~\bibnamefont
  {Stopa}}, \bibinfo {author} {\bibfnamefont {X.}~\bibnamefont {Andrade}},
  \bibinfo {author} {\bibfnamefont {M.~A.}\ \bibnamefont {Watson}}, \ and\
  \bibinfo {author} {\bibfnamefont {A.}~\bibnamefont {Aspuru-Guzik}},\
  }\href@noop {} {\bibfield  {journal} {\bibinfo  {journal} {J. Phys. Chem.
  Lett.},\ }\textbf {\bibinfo {volume} {2}},\ \bibinfo {pages} {682} (\bibinfo
  {year} {2011})}\BibitemShut {NoStop}%
\end{thebibliography}%

\end{document}
%